\documentclass[journal]{IEEEtran}
\IEEEoverridecommandlockouts
\usepackage{cite}
\usepackage{amsmath,amssymb,amsfonts}
\usepackage{algorithmic}
\usepackage{graphicx}
\usepackage{textcomp}
\usepackage{xcolor}
\usepackage{booktabs}
\usepackage{threeparttable}
\usepackage{multirow}
\usepackage{tabularx}
\usepackage{array}
\usepackage{lettrine}  
\usepackage{array}
\usepackage{nicefrac}

\newcolumntype{C}[1]{>{\centering\arraybackslash}m{#1}}

\def\BibTeX{{\rm B\kern-.05em{\sc i\kern-.025em b}\kern-.08em
    T\kern-.1667em\lower.7ex\hbox{E}\kern-.125emX}}

\usepackage{orcidlink}
\begin{document}

\title{A Reconfigurable Multiplier Architecture for Error-Resilient Applications in RISC-V Core}
\author{
\IEEEauthorblockN{
Pragun Jaswal\textsuperscript{*}~\orcidlink{0009-0005-9580-3184},
L.~Hemanth Krishna\textsuperscript{†}~\orcidlink{0000-0002-7737-6685},
B.~Srinivasu\textsuperscript{‡}~\orcidlink{0000-0003-0974-8245} \\
}
\IEEEauthorblockA{
Indian Institute of Technology Mandi, Mandi--175005, India\\
\textsuperscript{*}thepragun@gmail.com,
\textsuperscript{†}hemanth.krishna412@gmail.com,
\textsuperscript{‡}srinivasu@iitmandi.ac.in
}
}
\makeatletter
\def\@IEEEaftertitletext{\vspace{-1\baselineskip}}
\makeatother
\maketitle
\begin{abstract}

Neural Networks (NNs) have been widely adopted due to their outstanding efficacy and adaptability
across computer vision and deep learning applications. The optimization of NNs is necessary to enable
their deployment on energy constrained embedded devices, where the limited available energy poses a significant challenge 
for efficient inference.
This paper presents a runtime reconfigurable multiplier architecture integrated 
into the RISC-V core, targeting energy efficient neural network inference and edge AI applications. 
The proposed multiplier supports adaptability for exact and approximate computation with multiple 
configurable accuracy levels via a dedicated \textit{mulscr},
enabling fine-grained energy accuracy control within a standard processor pipeline. 
The proposed design achieves 44\%-52\% and 62\%-68\% power reduction in exact and approximate modes respectively, 
while maintaining the computational performance of 1.89 DMIPS/MHz. Evaluations on error-tolerant workloads 
including 2d convolution and matrix multiplication demonstrate up to 63\% reduction in energy consumption, 
with the proposed design achieving 1.21 pJ/instruction for matrix multiplication, confirming its effectiveness 
for energy-constrained edge AI deployments.

\end{abstract}


\section{Introduction}
\label{sec:intro}
Neural Networks (NNs) have achieved state-of-the-art performance in various application 
domains, including computer vision, natural language processing, and edge intelligence \cite{LeCun2015}. These 
networks mainly rely on highly parallel linear operations such as convolution and matrix multiplication, 
which dominate the computational workload. In recent years, the deployment of NNs in 
embedded systems has gained significant attention, as it enables intelligent processing directly at the edge, 
reducing latency, bandwidth usage, and dependence on cloud-based computation \cite{Li2020}. However, high accuracy of
NNs comes at the cost of high computational complexity, which makes their 
deployment challenging on power and resource limited  embedded devices \cite{Merenda2020}.
A notable observation is that NNs offers error tolerance capabilities.
Minor computation error in the arithmetic units  often do not noticeably 
degrade output quality \cite{Jiang2016}. This characteristic 
has encouraged the adoption of approximate computing, a design methodology that deliberately 
trades numerical precision for improvements in energy efficiency, latency, and silicon area.

Approximate computing reduces computational cost by allowing controlled inaccuracies in arithmetic operations 
\cite{Amirafshar2023,Guella2024,Anil2023,Kumari2025}.
However, the error sensitivity of neural networks varies across different layers and workloads \cite{Guella2024}, 
making a fixed-accuracy approximate unit inefficient. Therefore, designing a multiplier 
with runtime reconfiguration is important to ensure flexibility and adaptability.

The RISC-V Instruction Set Architecture (ISA) \cite{Cui2023} is now becoming a preferred choice among embedded 
processor developers, due to its adaptability and  flexibility for various applications. 
The integration of RISC-V with approximation computing \cite{Guella2024,Delavari2024DSD}, presents further opportunity to 
enhance the energy efficiency and performance of embedded processors. This work presents a novel hardware architecture 
for reconfigurable multiplier and also examines the effect of adaptive hardware approximation 
on the runtime energy efficiency of a RISC-V embedded processor for various applications.
The primary contributions of this study are as follows:
\begin{enumerate}
    \item Utilizing a 32-bit optimized 3-stage pipelined RV32I(E)M embedded processor using standard and special RISC-V Control Status Registers (CSRs).
    \item Design and implementation of an runtime reconfigurable unsigned multiplier with 255 approximation levels.
    \item Analysis and evaluation of runtime approximation at core level, in an embedded processor for error resilient applications.
    \item Proposed design in approximate mode achieves 1.21 pJ/instruction for matrix multiplication which is 67\% improvement from exact implementation.
\end{enumerate}

The rest of the paper is organized as follows: Section \ref{sec:Background} provides the background and related works of RISC-V based approximate computing. Section \ref{sec:hardware} describes the proposed hardware architecture, and Section \ref{sec:Implementation} presents its hardware implementation results and evaluation of the proposed architecture, followed by the conclusion in Section \ref{sec:conclusion}.

\section{Related Works}
\label{sec:Background}

Approximate computing has been explored across multiple levels of the hardware stack, 
from circuit level arithmetic units to processor micro-architectures. 
This section reviews the recent work presented in the literature.

\subsection{Circuit Level Approximate Multipliers}
At the circuit level, a substantial work has addressed the design of approximate 
adders and multipliers. Early works targeting image processing applications were reported in \cite{ Jiang2016}, 
with carry-disregard multipliers subsequently introduced in \cite{Amirafshar2023} to 
improve error characteristics alongside hardware complexity. Compressor based architectures have also been extensively studied,
including approximate 4:2 compressor based multipliers \cite{Strollo2020}, error recovery integrated design \cite{Anil2023},
and hybrid partial-product recursive multipliers \cite{Kumari2025}. 
Alternative strategies such as logarithmic and dynamic range multipliers 
targeting machine learning workloads have further been explored \cite{Yin2021}. 
However, the majority of these designs employ static approximation strategies, 
with fixed accuracy at design time and no mechanism for runtime control, limiting their 
applicability to dynamic and general-purpose computing scenarios.

\subsection{RISC-V Based Reconfigurable Approximate Computing}

The integration of approximate arithmetic into open ISAs such as RISC-V has recently 
attracted significant attention. ApproxRISC \cite{Trevisan2018} introduced one of the early designs enabling software-based approximate execution within a RISC-V framework,
while AxPIKE \cite{Felzmann2021} enabled instruction-level approximation evaluation 
for quality-energy trade-off analysis. Verma \textit{et al.} \cite{Verma2022} integrated a shift-and-add 
based approximate multiplier into a RI5CY core \cite{Davide2017} for error-tolerant applications. 
Delavari \textit{et al.} \cite{Delavari2024DSD} proposed the phoeniX platform, 
a modular RISC-V core supporting configurable approximate arithmetic units via CSR-based control, 
with dynamic switching between exact and approximate execution further evaluated in \cite{Delavari2024IICM}.
More recently, MARLIN \cite{Guella2024} proposed a runtime-reconfigurable signed multiplier
supporting 255 approximation levels integrated into a RI5CY-based core with CSR-controlled dynamic precision.

Moreover, most existing designs offer limited configurability and are tailored to specific application domains.
This work addresses these limitations by presenting a micro-architectural design of a runtime-reconfigurable
approximate multiplier that supports a wide and continuous range of accuracy levels, including exact computation, 
and is designed for seamless integration into a standard RISC-V processor pipeline.

\section{Proposed Reconfigurable Multiplier}
\label{sec:hardware}

\begin{figure}[h]
    \centering
    \includegraphics[width=0.7\linewidth]{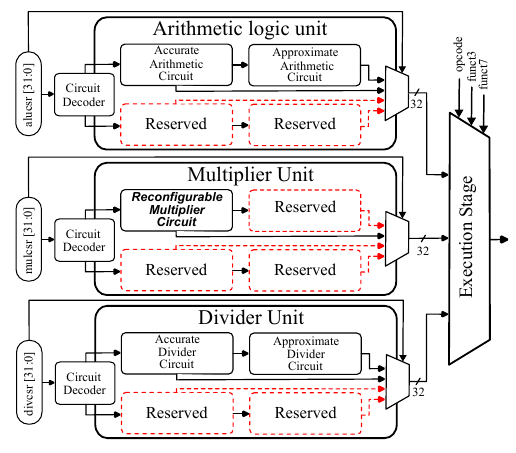}
    \caption{ Execution stage of the processor \cite{Delavari2024DSD} with proposed reconfigurable multiplier. }
    \label{fig:execution}
\end{figure}

The proposed design is implemented on the phoeniX processor \cite{Delavari2024DSD}, 
an open-source RISC-V core supporting the RV32I(E)M instruction set. 
phoeniX features a 3-stage scalar pipeline comprising an Instruction Fetch and Decode (IF/ID) stage, 
an Execution (EXE) stage, and a Memory Access with Write-back (MEM/WB) stage. 
The execution stage hosts three computational units, an Arithmetic Logic Unit, 
a Multiplier Unit, and a Divider Unit corresponding to the RV32IEM instruction set requirements. 
A key architectural feature of phoeniX is its support for runtime approximation control through 
three dedicated Control and Status Registers (CSRs) mapped at addresses \texttt{0x800} ($alucsr$), 
\texttt{0x801} ($mulcsr$), and \texttt{0x802} ($divcsr$) in the standard RISC-V CSR addressing space \cite{Cui2023}. 
These CSRs allow each execution unit to dynamically switch between exact and approximate operation, 
as well as control the level of approximation at runtime without any pipeline stall or structural modification.
Fig. \ref{fig:execution} illustrates the high level block diagram of the execution stage of the phoeniX core.
In the original phoeniX architecture, the multiplier unit consists of two separate circuits, one exact and one approximate.
In the proposed modification, both are replaced by a single proposed novel runtime-reconfigurable 32-bit multiplier that dynamically operates in either exact or approximate mode based on the \texttt{mulscr} configuration. 
This consolidation eliminates redundant hardware, reduces area overhead, and preserves full 
computational flexibility across all supported accuracy levels.

\begin{figure}[h]
    \centering
    \includegraphics[width=1\linewidth]{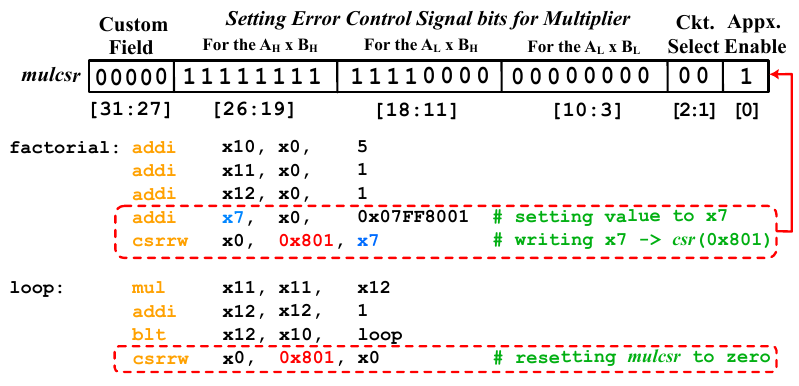}
    \caption{Sample assembly code for factorial using \texttt{mulscr} for approximation control.}
    \label{fig:assembly}
\end{figure}

The \texttt{mulscr} register provides fine-grained control over the approximation 
behavior of the multiplier unit, with each field governing a specific aspect of its operation. 
Fig. \ref{fig:assembly} presents a sample assembly code snippet illustrating the use 
of the CSR at address \texttt{0x801} to configure the \texttt{mulscr} fields.
The $mulcsr[0]$ serves as the approximation enable flag, `1' activates approximate computation, else exact multiplication. 
The $mulcsr[2:1]$ in the original phoeniX are for circuit selection, controlling the multiplexer. While the proposed design integrates a single reconfigurable multiplier, these bits set to \texttt{`00'}; however, they are retained for compatibility with future multi unit configurations. The 8-bit error control for 8-bit multiplier to provide 255 approximation levels is through $mulcsr[26:3]$. The 
$mulcsr[10:3]$ controls the approximation level of the lower 8-bit multiplier $A_L \times B_L$. While $mulcsr[18:11]$ are for multipliers $A_L \times B_H$ and $A_H \times B_L$, and $mulcsr[26:19]$
for multiplier $A_H \times B_H$.
This multiplier configuration is further discussed later in this section. 
Finally, $mulcsr[31:27]$ are reserved for custom fields, available for application specific control logic and future extensions.

\subsection{Reconfigurable 4:2 Compressor }
\begin{figure}[h]
    \centering
    \includegraphics[width=1\linewidth]{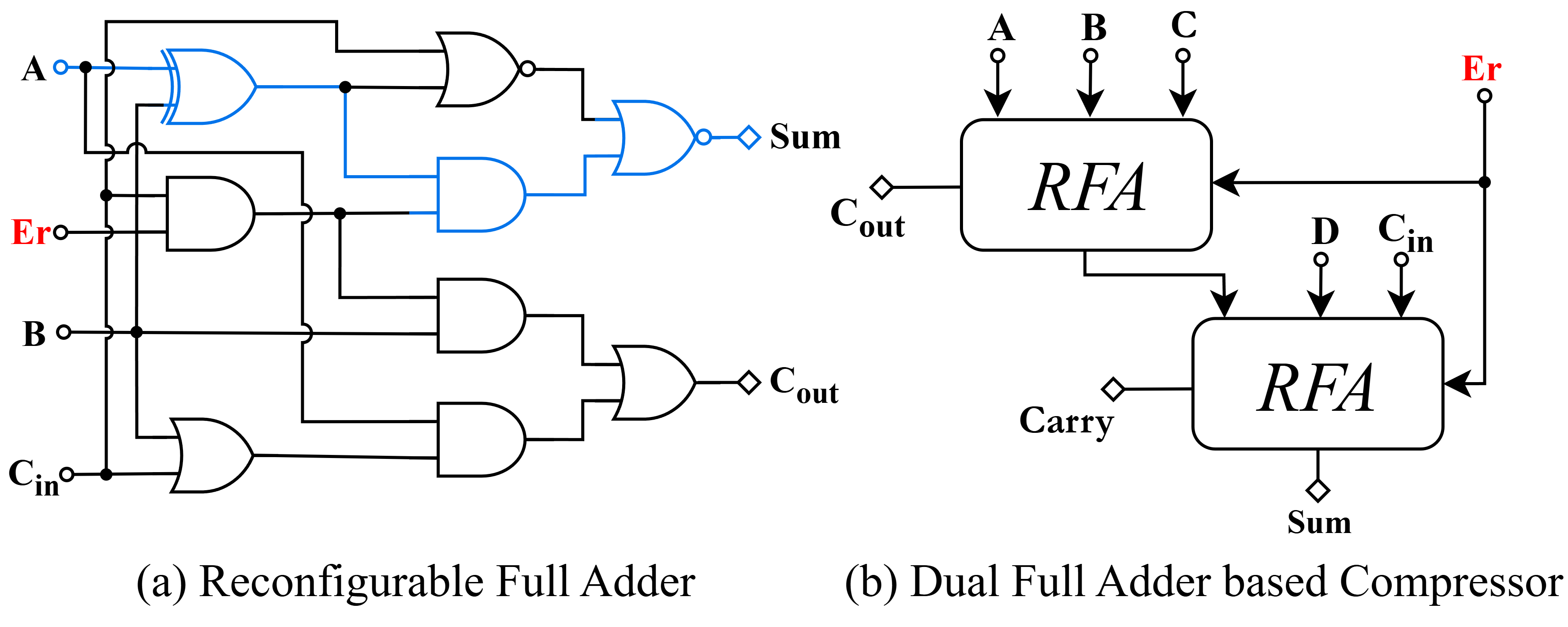}
    \caption{Proposed Reconfigurable Designs: (a) \textit{RFA}, and (b) \textit{DFC}.}
    \label{fig:Full_adder}
\end{figure}

\begin{figure}[h]
    \centering
    \includegraphics[width=0.6\linewidth]{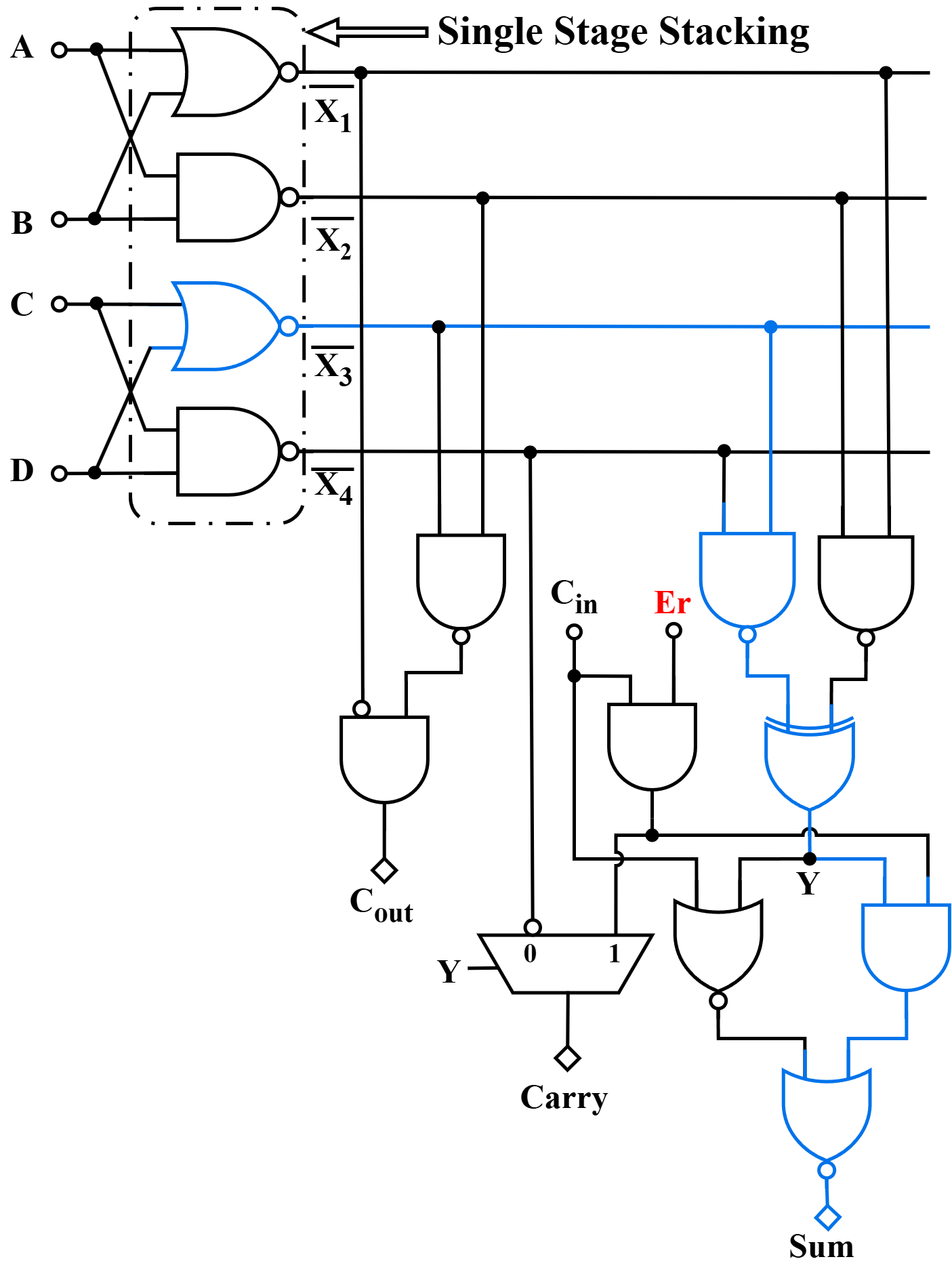}
    \caption{Proposed Single Stacking-based reconfigurable Compressor \textit{(SSC)}.}
    \label{fig:Reconfigurable_compressor}
\end{figure}

Two designs of a reconfigurable 4:2 compressor are proposed. 
The first design is constructed reconfigurable full adder (RFA), 
as illustrated in Fig. \ref{fig:Full_adder} (a). The \textit{RFA} can operate either 
exact or approximate mode 
depending on the value of the error signal (Er). When 
Er = `1', the circuit functions as an exact full adder, whereas Er = `0' enables 
approximate computation with reduced switching activity.
Using two such \textit{RFAs}, we propose a dual full adder based reconfigurable 4:2 compressor \textit{(DFC)}, as depicted in Fig. \ref{fig:Full_adder} (b).
The second design directly implements with the help of single stage stacking logic \cite{anurag2025}.
The proposed single stacking based reconfigurable 4:2 compressor \textit{(SSC)}
is shown in Fig. \ref{fig:Reconfigurable_compressor}.
The proposed \textit{DFC} design generates 13 erroneous output combinations
out of 32 possible input cases, with an Error Distance (ED) of $\pm1$ and $-2$.
In contrast, the \textit{SSC} design exhibits only 8 erroneous cases
with a maximum error distance of $+1$, as presented in Table \ref{tab:compressor_truth}.

\begin{table}[!t]
\caption{Truth table of the proposed reconfigurable 4:2 compressor}
\label{tab:compressor_truth}
\centering
\setlength{\tabcolsep}{3pt}
\resizebox{\linewidth}{!}{
\begin{tabular}{|c|c|c|c|c|c|c|c|c|c|c|c|c|}
\hline
\multicolumn{5}{|c|}{Inputs} & \multicolumn{4}{c|}{\textit{DFC}}  & \multicolumn{4}{c|}{\textit{SSC}} \\ \hline
$X_1$ & $X_2$ & $X_3$ & $X_4$ & $C_{in}$ & $C_{out}$ & Carry & Sum & ED &$C_{out}$ & Carry & Sum & ED\\ \hline
0 & 0 & 0 & 1 & 1   & 0 & 1 & 1 & +1  & 0 & 1 & 1 & +1\\
0 & 0 & 1 & 0 & 1   & 0 & 0 & 1 & -1  & 0 & 1 & 1 & +1\\
0 & 1 & 0 & 0 & 1   & 0 & 0 & 1 & -1  & 0 & 1 & 1 & +1\\

0 & 1 & 1 & 0 & 0   & 0 & 0 & 1 & -1  & 0 & 1 & 0 & 0\\
0 & 1 & 1 & 0 & 1   & 0 & 0 & 1 & -2  & 0 & 1 & 1 & 0\\
0 & 1 & 1 & 1 & 0   & 0 & 1 & 0 & -1  & 0 & 1 & 1 & 0\\

0 & 1 & 1 & 1 & 1   & 0 & 1 & 1 & -1  & 1 & 1 & 1 & +1\\
1 & 0 & 0 & 0 & 1   & 0 & 0 & 1 & -1  & 0 & 1 & 1 & +1\\
1 & 0 & 1 & 0 & 0   & 1 & 0 & 1 & +1  & 0 & 1 & 0 & 0\\

1 & 0 & 1 & 1 & 0   & 1 & 1 & 1 & +1  & 0 & 1 & 1 &  0\\
1 & 0 & 1 & 1 & 1   & 1 & 1 & 1 & +1  & 1 & 1 & 1 & +1\\
1 & 1 & 0 & 1 & 1   & 1 & 1 & 1 & +1  & 1 & 1 & 1 & +1\\
1 & 1 & 1 & 0 & 1   & 1 & 0 & 1 & -1  & 1 & 1 & 1 & +1\\

\hline
\end{tabular}
}
\end{table}





\subsection{Reconfigurable Multi-bit Multiplier}
\begin{figure}[h]
    \centering
    \includegraphics[width=0.7\linewidth]{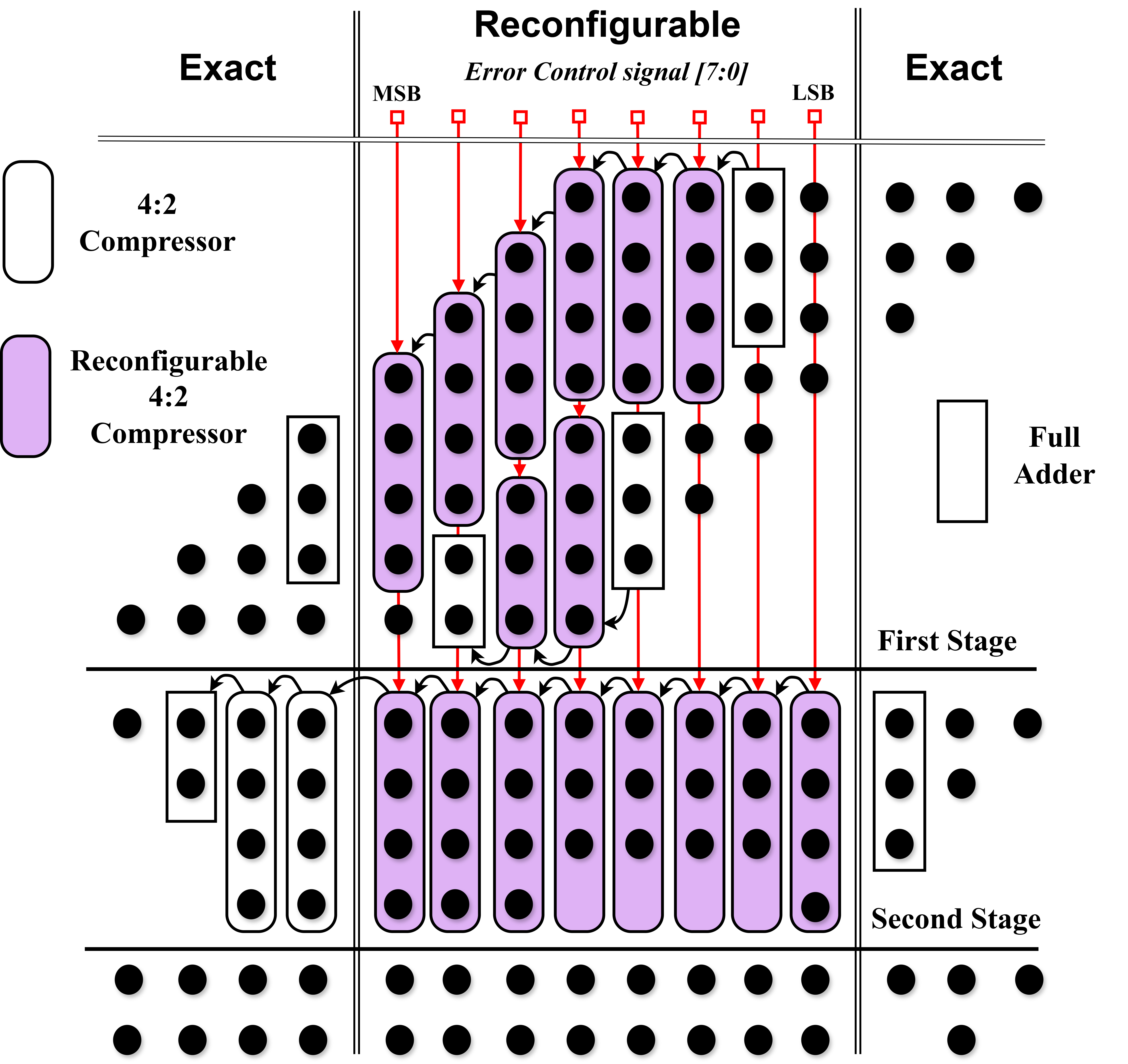}
    \caption{Proposed Reconfigurable multiplier with control signal ER.}
    \label{fig:Reconfigurable_multiplier}
\end{figure}

Fig. \ref{fig:Reconfigurable_multiplier} illustrates the dot diagram of the proposed
8-bit unsigned multiplier. 
Columns [11:4] are designated as the reconfigurable region, which is controlled by an 8-bit error control signal (Er) 
to dynamically adjust the approximation level at runtime. 
Based on the proposed reconfigurable compressors, two 8-bit multiplier variants are designed,
the Dual Full adder based Multiplier (\textit{DFM}), which employs DFC compressors in the reconfigurable region, 
and the Single Stacking based Multiplier (\textit{SSM}), which utilizes SSC compressors in the reconfigurable region.

\begin{figure}[h]
    \centering
    \includegraphics[width=1\linewidth]{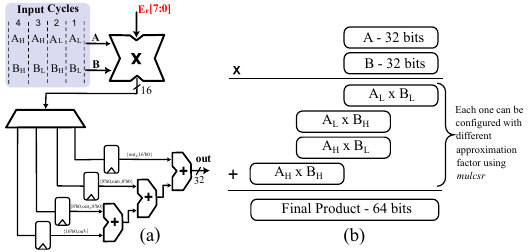}
    \caption{Reconfigurable multiplier: (a) 16-bit, and (b) 32-bit.}
    \label{fig:final_mul32}
\end{figure}

Fig. \ref{fig:final_mul32} illustrate the 16-bit and 32-bit multipliers using
the proposed 8-bit multiplier.
The 16-bit multiplier presented in Fig. \ref{fig:final_mul32} (a)
which uses a single 8-bit multiplier, is reused over four consecutive cycles to perform a 16-bit multiplication, 
minimizing hardware resources while maintaining correctness. 
This 16-bit structure is then replicated four times, as shown in Fig. \ref{fig:final_mul32} (b), 
to realize the full 32-bit multiplication required by the RV32IEM instruction set. 
A notable advantage of this hierarchical configuration is that 
each 16-bit multiplier unit can be independently configured to a different approximation level via the \texttt{mulscr}, 
enabling fine-grained accuracy control across the 32-bit datapath.

\section{Evaluation and Applications}
\label{sec:Implementation}
In this section, we analyze the performance of the proposed
designs using the FPGA and ASIC synthesis. Furthermore, the assessment
is conducted across various applications in RISC-V core to demonstrate 
the effectiveness of the design.

\subsection{Hardware Evaluation}

All presented designs are implemented in Verilog HDL and synthesized using Cadence Genus Synthesis Solution 
targeting $UMC$ $90nm$ technology under typical-typical (TT) process conditions at 25°C. 
Post-synthesis evaluation is carried out based on critical-path delay, power consumption, and area. 
Value Change Dump (VCD) files are generated across different approximation levels to capture 
realistic switching activity for accurate power estimation.

Table \ref{tab:compressor_comparison} summarizes the error, area, power, and delay characteristics of the proposed 4:2 compressors. 
Since no reconfigurable compressor designs are reported in the literature, both proposed designs,
\textit{DFC} and \textit{SSC} are evaluated against their exact implementation. 
The \textit{DFC} design exhibits an Error Rate (ER) of $\nicefrac{13}{32}$ while achieving an energy improvement of 10\% in approximate mode. 
The \textit{SSC} design, offering a lower ER of $\nicefrac{8}{32}$, achieves an energy improvement of 8.6\%.
These results indicate that \textit{DFC} offers better energy savings at the cost of higher approximation error, 
while \textit{SSC} provides a more conservative accuracy-efficiency trade-off.

\begin{table}[h]
\centering
\caption{Hardware Analysis of 4:2 Compressors}
\label{tab:compressor_comparison}

\renewcommand{\arraystretch}{1.2}
\setlength{\tabcolsep}{7pt}

\begin{tabular}{|c|c|c|c|c|c|}
\hline
\multirow{2}{*}{\textbf{Design}} & \multirow{2}{*}{\textbf{Error}} & \textbf{Area} & \textbf{Power} & \textbf{Delay} & \textbf{Energy} \\
 &  & ($\mu m^2$) & ($\mu W$) & (ps) & (aJ) \\
\hline
Exact    & -                   & 45.47 & 6.12        & 296 & 1811 \\
\hline
\multirow{2}{*}{\textit{DFC}} & \multirow{2}{*}{$\nicefrac{13}{32}$} & \multirow{2}{*}{57.23} & 5.84 - & \multirow{2}{*}{279} & 1629 - \\
  &  &                                   & 8.01  &     & 2236 \\
\hline
\multirow{2}{*}{\textit{SSC}} & \multirow{2}{*}{$\nicefrac{8}{32}$}  & \multirow{2}{*}{79.39} & 6.84 -   & \multirow{2}{*}{242} & 1655 - \\
 &   &  & 7.89   &  & 1909 \\
\hline
\end{tabular}

\end{table}

Table \ref{tab:Multiplier_comparison} summarizes the error, area, power, and delay characteristics of the different 8-bit unsigned multipliers. Existing approximate designs \cite{Amirafshar2023,Anil2023,Kumari2025} achieve notable energy reductions but with fixed error rates, limiting their use in general-purpose computing, while the reconfigurable design in \cite{Delavari2024DSD} supports configurability but lacks exact computation mode. The proposed \textit{DFM} and \textit{SSM} multipliers overcome both limitations by supporting exact and approximate modes with runtime-configurable accuracy levels via \texttt{mulcsr}. In approximate mode, DFM achieves 28\% energy savings with an error rate of 75.7\% compared to the exact Dadda multiplier, while SSM achieves 23\% energy savings at a lower error rate, making it preferable for applications requiring hard error bounds.

\begin{table}[]
\centering
\caption{Hardware Analysis of 8 bit Unsigned Multiplier}
\label{tab:Multiplier_comparison}
\renewcommand{\arraystretch}{1.3}
\setlength{\tabcolsep}{5pt}
\begin{tabular}{ccccccc}
\toprule
\multirow{2}{*}{\textbf{Design}}  & \textbf{Area} & \textbf{Power} & \textbf{Delay} & \textbf{Energy} & \textbf{ER} & \textbf{MRED}\\
 &  ($\mu m^2$) & ($\mu W$) & (ns) & (pJ) & (\%) & (\%) \\
\midrule

Dadda Exact            & 1360.10 & 257.19  & 1.50 & 385.7 & 0 &  0 \\
\midrule

\multicolumn{7}{c}{\textit{Fixed Approximate Designs}} \\
\midrule
\cite{Amirafshar2023}  & 1339.7 & 93.7  & 2.43 & 227.6 & NR* & 3.46  \\
\hline
\cite{Anil2023}        & 987.7  & 135.1 & 2.07 & 279.6 & NR* & 0.715 \\
\hline
\cite{Kumari2025}      & 1021.10 & 115.3 & 1.99 & 229.4 & 84 & 5.33  \\
\midrule

\multicolumn{7}{c}{\textit{Reconfigurable Designs}} \\
\midrule

\multirow{2}{*}{\cite{Delavari2024IICM}} & \multirow{2}{*}{860} & 178 - & \multirow{2}{*}{1.34} & 238 - & 36.16 - & 0.85 - \\
                            &                      & 196  &                      & 262  & 65.06  & 8.94 \\
\hline

\multirow{2}{*}{\textit{DFM}} & \multirow{2}{*}{1419.2} & 196 - & \multirow{2}{*}{1.42} & 278 - & 0 - & 0 - \\
                          &                      & 355  &                      & 504  & 75.70 & 5.89\\
\hline

\multirow{2}{*}{\textit{SSM}} & \multirow{2}{*}{1319.4} & 231 - & \multirow{2}{*}{1.28} & 295 - & 0 - & 0 - \\
                          &                     & 315  &                      & 403  & 66.65 & 7.68 \\
\bottomrule
\end{tabular}

\begin{flushleft}
\small
\ \ \ \ NR* : Not Reported.
\end{flushleft}

\end{table}

\begin{figure}
    \centering
    \includegraphics[width=1\linewidth]{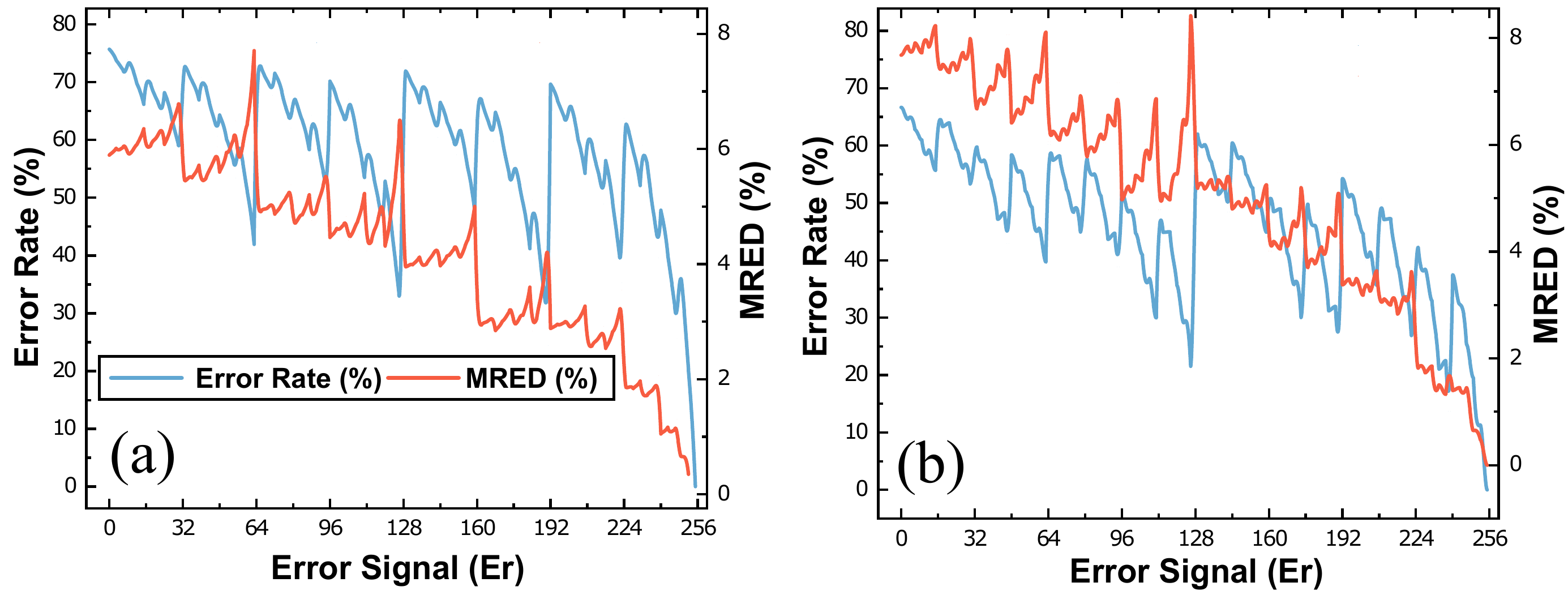}
\caption{Variation of MRED and error rate in percentage with error control signal for the proposed 8-bit multipliers: (a) \textit{DFM} and (b) \textit{SSM}. 
    Where $Er = 0$ represents maximum approximation and $Er = 255$ represents exact mode.}
    \label{fig:Error}
\end{figure}

Fig. \ref{fig:Error} illustrates the variation of Mean Relative Error Distance (MRED) and error rate 
across all approximation levels for both proposed 8 bit multiplier designs. The jumps and drops in MRED and error rate, 
respectively can be understood by referring to the multiplier structure in Fig. \ref{fig:Reconfigurable_multiplier}.
It can be observed that \textit{SSM} exhibits a lower error rate but higher MRED compared to DFM. 
This behavior is attributed to the underlying compressor designs, \textit{SSM} employs the SSC compressor, 
which has less wrong combinations but a consistently positive error of $+1$, resulting in a lower error rate but an accumulated positive MRED.
In contrast, \textit{DFM} employs the \textit{DFC} compressor, which introduces both positive and negative errors of $\pm1$ and $-2$,
leading to a higher error rate but partial error cancellation moderates the overall MRED.
Since partial products contribute unequally to the final result depending on their column position due to different weights, 
approximating higher significance columns produces larger jumps in MRED compared to lower significance ones. 
This explains the sharp increases in MRED observed at configuration boundaries such as 63→64 and 127→128, 
where the approximation transitions to a more significant column. Conversely, the error rate decreases at higher approximation factors, 
as fewer bit positions contribute distinct error events at those stages of the computation.

Table \ref{tab:embedded_processors} presents a comparison of the proposed design 
against representative embedded cores, evaluated in terms of LUT utilization on a Xilinx 7-series FPGA, ASIC parameters
and computational performance in Dhrystone Million Instructions Per Second per Megahertz (DMIPS/MHz). 
The compared cores include RI5CY \cite{Davide2017} and CV32E40P \cite{Noam2021}, 
both featuring a 4-stage in-order pipeline implemented in 65nm CMOS, Ibex \cite{Noam2021},
an area-optimized 2-stage pipeline core achieving 0.9 DMIPS/MHz, and PicoRV32 \cite{Djupdal2023}, a lightweight 
open-source core targeting minimal footprint ASIC implementations. The original phoeniX core 
is reproduced under UMC 90nm as a baseline, with two multiplier units (exact and approximate) and default ALU and divider configurations. 
The proposed modification, which consolidates both multiplier units into a 
single runtime-reconfigurable unit, achieves 13\% reduction in area and 11\% reduction in power consumption 
while maintaining the same performance of 1.89 DMIPS/MHz, 
demonstrating the best area-power-performance trade off among all compared cores.

\begin{table}
\centering
\caption{Architectural and Performance Comparison of Embedded Class Processors}
\label{tab:embedded_processors}

\renewcommand{\arraystretch}{1.5}
\setlength{\tabcolsep}{3pt}
\resizebox{\linewidth}{!}{%
\begin{tabular}{|c|c|c|c|c|c|c|c|}
\hline
\multirow{2}{*}{\textbf{Processor}} & \multirow{2}{*}{\textbf{LUTs}} & \textbf{Node} & \textbf{Power}  & \textbf{Freq.} & \textbf{Area } & \textbf{DMIPS} & \multirow{2}{*}{\textbf{ISA}} \\
 &  & (nm) & (mW)  & (MHz) & (mm$^2$) & \textbf{/MHz} &  \\
\hline
RI5CY \cite{Davide2017} & NR* & 65 & 3.77 & 560 & 0.059 & 1.10 & RV32IM  \\
\hline
CV32E40P \cite{Noam2021} & NR* & 65 & 1.57  & 453 & 0.104 & NR* & RV32IMC  \\
\hline
Ibex \cite{Noam2021} & NR* & 65 & 1.78  & 394 & 0.053 & 0.90 & RV32IMC  \\
\hline
PicoRV32 \cite{Djupdal2023} & 1765 & 130 & 5.14  & 250 & 0.23 & 0.50 & RV32IMC  \\
\hline
phoeniX \cite{Delavari2024DSD} & 4552 & 90 & 60.26  & 620 & 0.110 & 1.89 & RV32IEM  \\
\hline
Proposed & 4365        & 90 & 53.68  & 620 & 0.0961 & 1.89 & RV32IEM  \\
\hline

\end{tabular}}
\begin{flushleft}
\small
\ \ \ \ NR* : Not Reported.
\end{flushleft}

\end{table}

\begin{figure}[]
    \centering
    \includegraphics[width=1\linewidth]{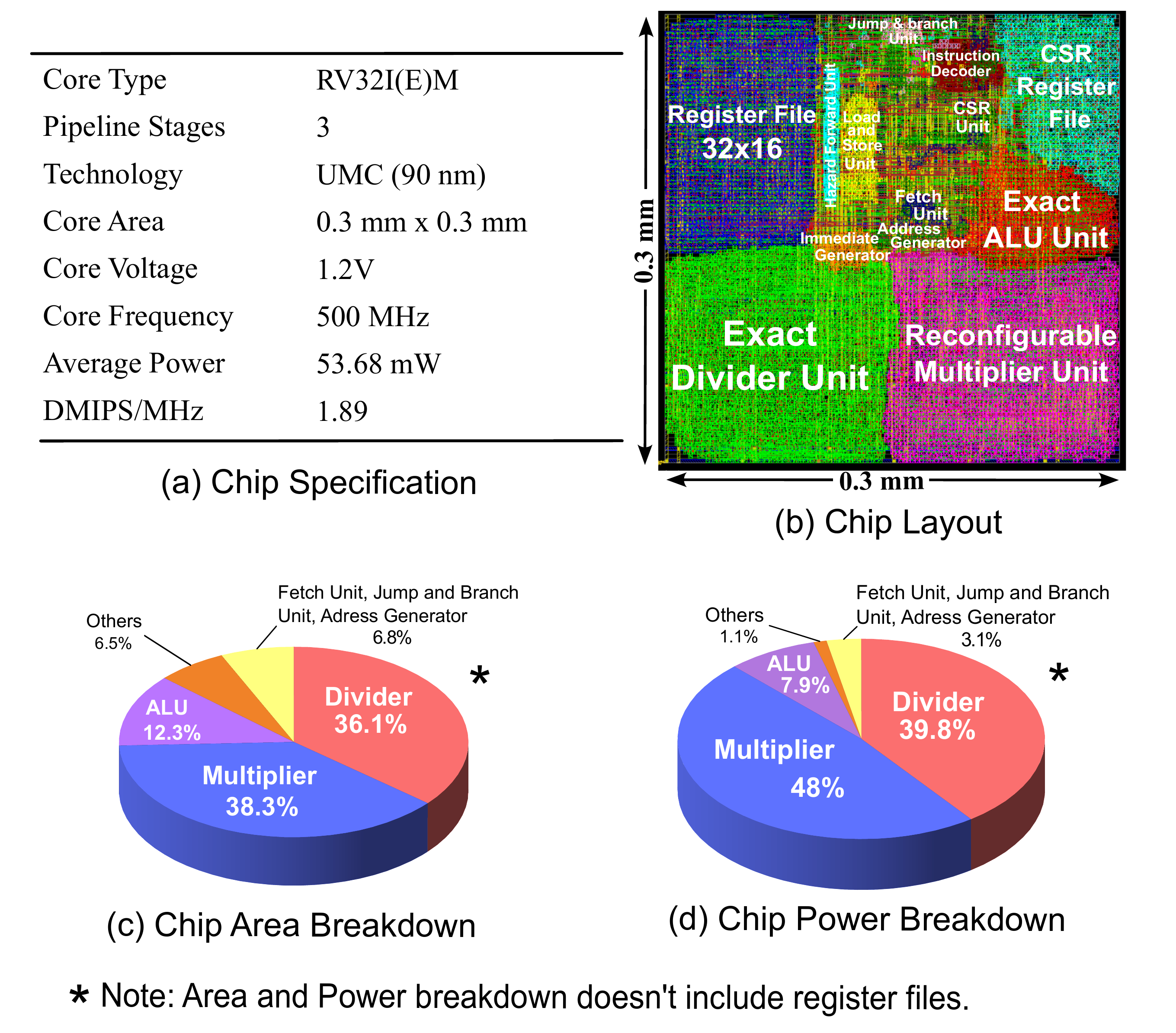}
    \caption{PhoeniX processor with modified execution unit (a) core specifications, (b) chip layout, (c) area distribution, and (d) power distribution.}
    \label{fig:evaluation_1}
\end{figure}

Fig. \ref{fig:evaluation_1} (a) and (b) shows the core specifications and chip layout having the core area of $0.3mm \times0.3mm$.
Fig. \ref{fig:evaluation_1} (c) and (d) presents for an estimation of the are and power breakdown in
the core, excluding memories and register files.
It can be concluded from Fig. \ref{fig:evaluation_1} (c) that the execution stage has
the highest area and power consumption in the inner core micro-architecture.
In the target core, the execution stage, including the
ALU, multiplier, and divider units, 
have the largest share of the area and power distribution
chart, accounting for 86.7\% and 95.7\%, respectively. The results also shows that 
only multiplier contributes for 48\% of power consumption,
whereas the multiplier in \cite{Delavari2024DSD} reports 
53\% in the overall power consumption of the 
RISC-V core with two multipliers. 

\subsection{Applications}
\begin{figure}[]
    \centering
    \includegraphics[width=0.9\linewidth]{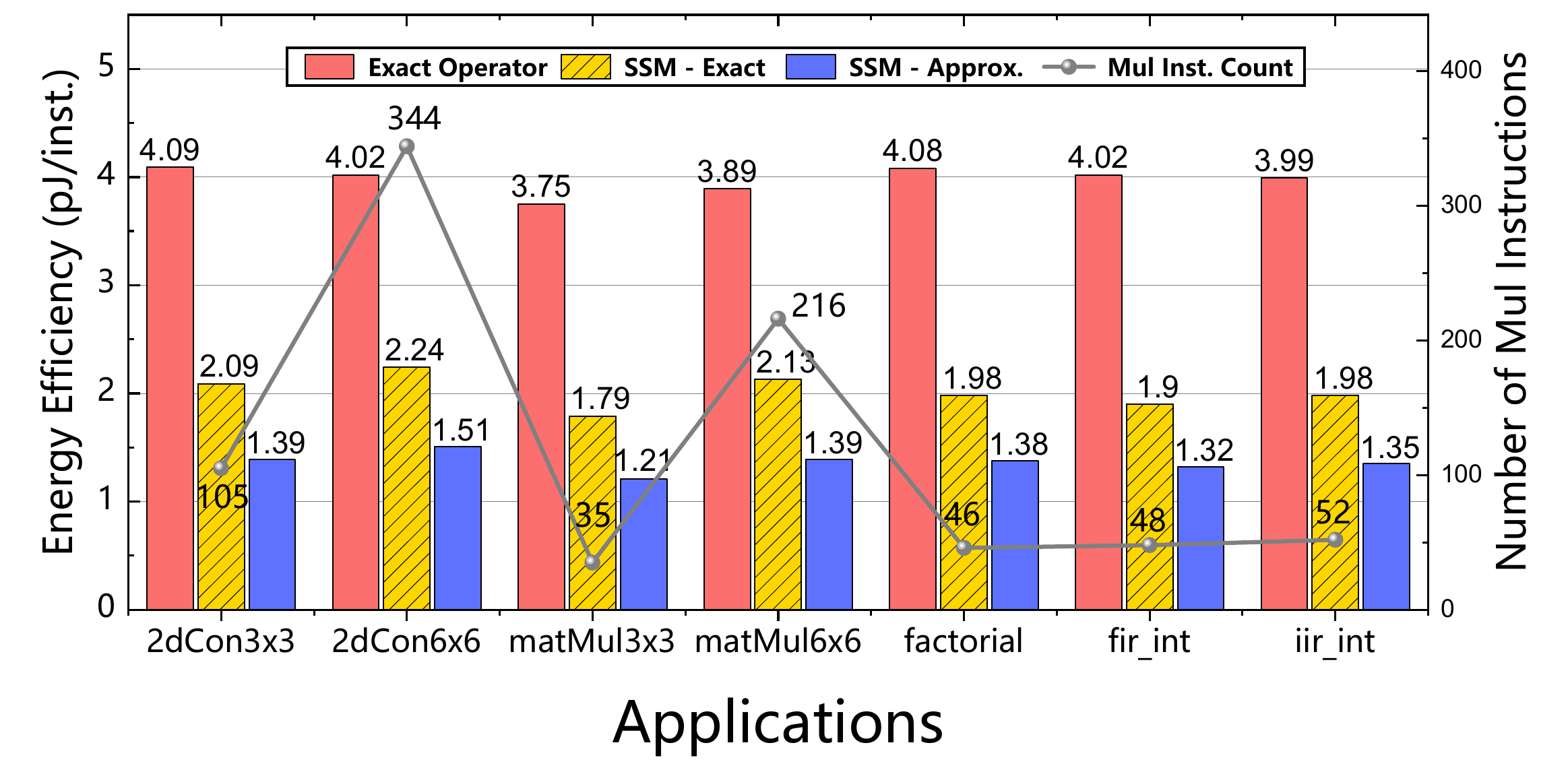}
    \caption{Energy efficiency in \textit{pJ/instruction} and multiplication instruction count (mul and mulh) across different benchmark applications.}
    \label{fig:Energy_efficiency}
\end{figure}

To validate the proposed design, evaluations are conducted on a set of standard
programs with a significant multiplication load. The selected applications include
2d convolution for CNN workloads and matrix multiplication for Transformer,
both of which are inherently error-tolerant and well-suited for approximate computing. 
All programs are compiled using the RISC-V GCC toolchain (riscv64-unknown-gcc 8.3.0) \cite{riscv_gnu_toolchain} 
with consistent compilation options across all experiments. Proposed \textit{SSM} multiplier design is used in 
exact and approximate mode with having \texttt{mulscr} value 0x00000000 and 0x00000001, respectively.
Fig. \ref{fig:Energy_efficiency} presents the energy efficiency in \textit{pJ/instruction} for each program,
evaluated under three configurations: the exact multiplier (using multiplication operator), and the proposed multiplier in both
exact and approximate modes. Execution time and instruction count are measured using standard RISC-V CSRs. Among all benchmarks,
matrix multiplication ($3\times3$) benefits the most from approximation, 
with the proposed multiplier achieving up to 63\% reduction in energy consumption in approximate mode.

\begin{table}[]
\centering
\caption{Power Consumption of Different Multiplier Units in RISC-V }
\label{tab:power_mul_units}
\renewcommand{\arraystretch}{1.2}
\setlength{\tabcolsep}{10pt}

\begin{tabular}{|c|c|c|c|c|}
\hline
\multirow{2}{*}{Applications} & \multirow{2}{*}{CPI} & 
\multicolumn{3}{c|}{Power Consumption (mW)} \\ \cline{3-5}

&& Exact & \textit{SSM-E} & \textit{SSM-A} \\

\hline
2dConv3x3 & 1.35  & 1.508 & 0.772 & 0.514  \\ \hline
2dConv6x6 & 1.37  & 1.462 & 0.814 & 0.551  \\ \hline
matMul3x3 & 1.29  & 1.450 & 0.692 & 0.467  \\ \hline
matMul6x6 & 1.34  & 1.452 & 0.795 & 0.521  \\ \hline
factorial & 1.39  & 1.460 & 0.710 & 0.497  \\ \hline
fir\_int  & 1.30  & 1.529 & 0.755 & 0.502  \\ \hline
iir\_int  & 1.31  & 1.509 & 0.751 & 0.511  \\ \hline

\end{tabular}
\end{table}

\begin{figure}[]
    \centering
    \includegraphics[width=0.9\linewidth]{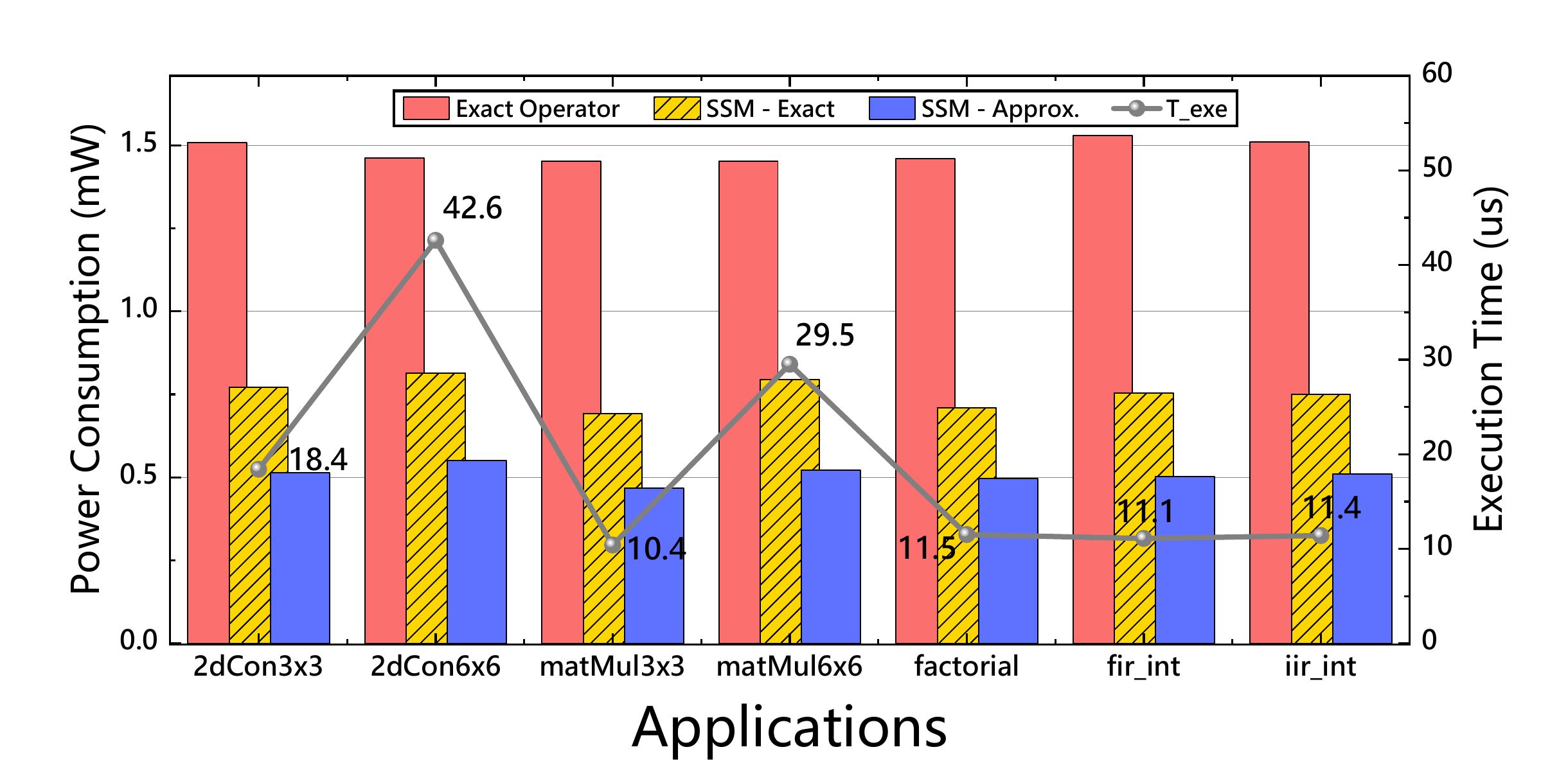}
    \caption{Power consumption of multiplier unit with overall execution time.}
    \label{fig:Power_consumption}
\end{figure}

\begin{figure}[]
    \centering
    \includegraphics[width=0.9\linewidth]{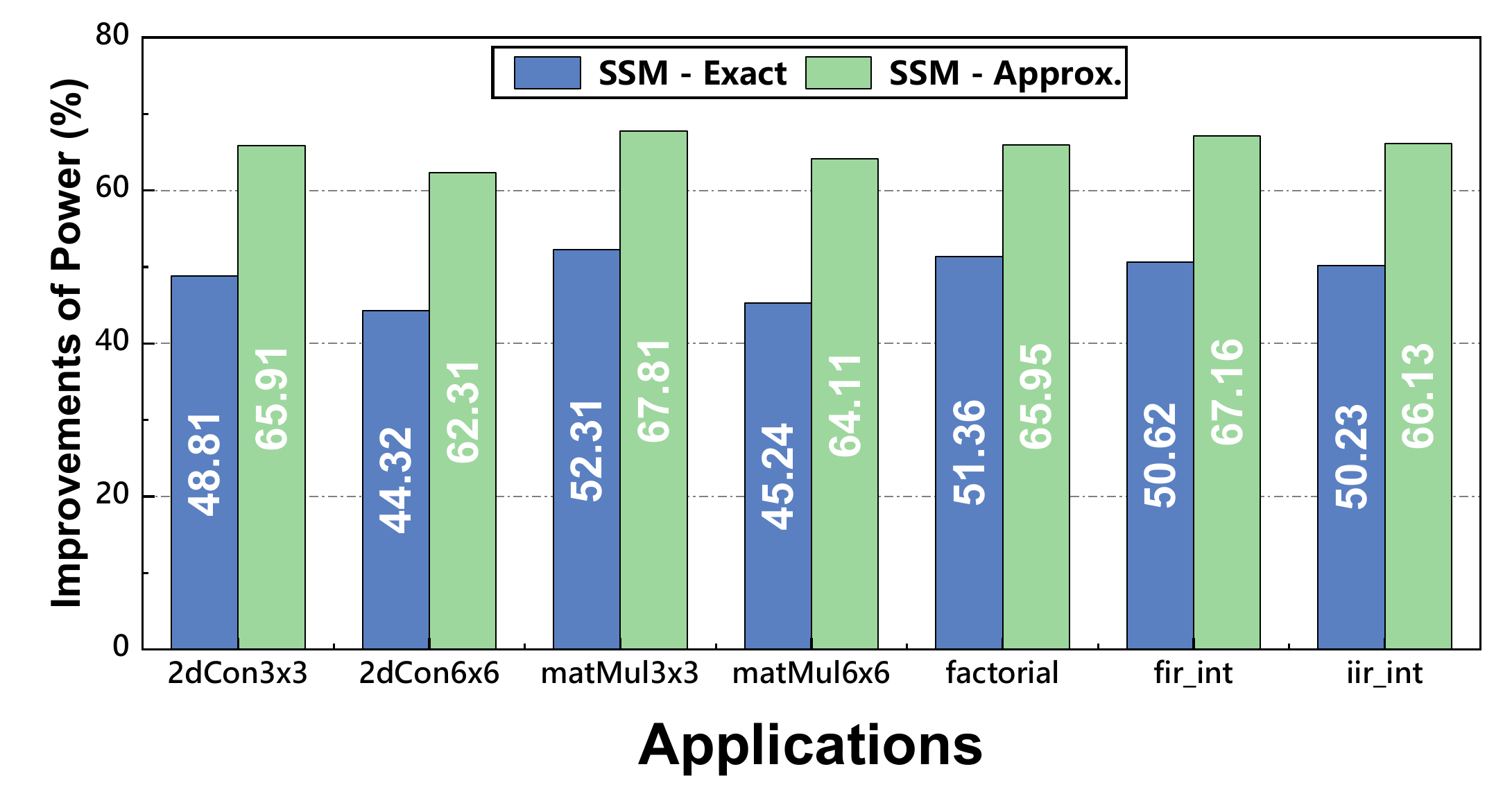}
    \caption{Improvements in power consumption for SSM in exact and approximate mode.}
    \label{fig:improvements}
\end{figure}

Table \ref{tab:power_mul_units} presents the power consumption and Cycles Per Instruction (CPI) values of the exact multiplier and the proposed \textit{SSM} operating in both exact and approximate modes across all benchmark applications. Fig. \ref{fig:Power_consumption} further illustrates the power consumption of different multiplier configurations across six applications, with the corresponding execution time in a gray line. As summarized in Fig. \ref{fig:improvements}, the proposed \textit{SSM} in exact mode achieves 44\% to 52\% reduction in power consumption, while in approximate mode the savings increase to 62\%–68\%, demonstrating consistent and significant power efficiency gains across all evaluated workloads.

\section{Conclusion}
\label{sec:conclusion}

This work presented a runtime reconfigurable multiplier architecture integrated into the phoeniX RISC-V core for energy-efficient edge AI and neural network inference applications. Two reconfigurable 4:2 compressor designs, \textit{DFC} and \textit{SSC}, are proposed to construct the \textit{DFM} and \textit{SSM} multipliers, supporting exact and approximate modes through runtime control using \texttt{mulcsr}. By replacing two conventional multiplier units with a single reconfigurable unit, the design achieved 13\% area reduction and 11\% lower core power consumption, demonstrating an efficient and scalable solution for runtime adaptive multiplier for edge AI devices.

\bibliographystyle{IEEEtran}
\bibliography{IEEEabrv,main}

\end{document}